\documentclass[12pt,aps,tightenlines,notitlepage,showpacs,superscriptaddress]{revtex4-1}
\usepackage[dvips]{epsfig}

\usepackage{amssymb}
\usepackage{amsfonts}
\usepackage{amsmath}
\usepackage{bm}
\usepackage{framed}
\usepackage{graphicx}
\usepackage{subfigure}
\usepackage{euscript}
\usepackage{wasysym}
\usepackage[margin=2.5cm]{geometry}
\usepackage{fancyhdr}
\usepackage{enumerate}
\usepackage{color}
\usepackage{float}
\usepackage{marvosym}
\usepackage{url}
\usepackage{enumitem} 

\usepackage{tikz}
\newcommand{\stencilpt}[4][]{\node[circle,draw,inner sep=0.1em,minimum size=0.8cm,font=\tiny,#1] at (#2) (#3) {#4}}

\usepackage{xparse}

\usepackage{xcolor}
\usepackage{listings}
\NewDocumentCommand{\codeword}{v}{%
\texttt{\textcolor{black}{#1}}%
}

\usepackage{lineno}

\newcommand{\mathd}{\mathrm{d}}

\newcommand{\mydim}{D}

\newcommand{\vecx}{\mathbf{x}}

\begin{document}

\title{cuSten -- CUDA Finite Difference and Stencil Library}

\author{Andrew Gloster}
\affiliation{School of Mathematics and Statistics, University  College Dublin, Belfield, Dublin 4, Ireland}
\author{Lennon \'O N\'araigh}
\affiliation{School of Mathematics and Statistics, University  College Dublin, Belfield, Dublin 4, Ireland}
\date{\today}


\begin{abstract}
In this paper we present cuSten, a new library of functions to handle the implementation of 2D and batched 1D finite-difference/stencil programs in CUDA. 
cuSten wraps data handling, kernel calls and streaming into four easy to use functions that speed up development of numerical codes on GPU platforms. 
The paper also presents an example of this library applied to solve the Cahn-Hilliard equation utilizing an ADI method with periodic boundary conditions, this solver is also used to benchmark the cuSten library performance against a serial implementation.

\end{abstract}


\maketitle




\section*{Current Software Version}
\label{}
\begin{table}[!h]
\begin{tabular}{|l|p{6.5cm}|p{8.5cm}|}
\hline
\textbf{Nr.} & \textbf{Code metadata description} & \\
\hline
C1 & Current code version & 2.1 \\
\hline
C2 & Permanent link to code/repository used of this code version & \url{https://github.com/munstermonster/cuSten/releases/tag/2.1} \\
\hline
C3 & Legal Code License   & Apache License 2.0 \\
\hline
C4 & Code versioning system used & git \\
\hline
C5 & Software code languages, tools, and services used & CUDA, C++, HDF5 (for one of the examples, not needed to compile library) \\
\hline
C6 & Compilation requirements, operating environments \& dependencies & CUDA\\
\hline
C7 & If available Link to developer documentation/manual & Generated using Makefile supplied with software, see README \\
\hline
C8 & Support email for questions & andrew.gloster@ucdconnect.ie\\
\hline
\end{tabular}
\caption{Code metadata}
\label{} 
\end{table}
Keywords: CUDA, Finite Difference, Library, PDEs, Stencil, Benchmark


\section{Introduction}
\label{sec:intro}
Many problems in Physics and Applied Mathematics can be expressed as systems of Partial Differential Equations (PDEs), examples of which include the Navier--Stokes \cite{doering1995applied}, Euler \cite{osher2006level, hesthaven2018numerical}, Black--Scholes \cite{wilmott_howison_dewynne_1995} and Burgers \cite{whitham2011linear} equations. 
In many situations analytic methods of solving a given system are not possible due to the complexity of the equations; an alternate approach is to solve the system numerically.
To discretize the system numerically, several standard approaches exist, including the finite-difference, finite-volume, and finite-element methods.   For definiteness, this work focus on the finite-difference method, however, it can be applied in any situation requiring stencil-based operations.

The application of the finite-difference method turns the operators in PDEs into expressions which can  be input into a computer program. 
For high-resolution numerical simulations, numerical scheme allowing, it is desirable to solve these computational problems in parallel with multiple processors to reduce the time taken to find a solution, this has been traditionally tackled with the MPI or OpenMP libraries which allow for parallelisation across multiple CPU cores.
More recently due to developments in technology, a reduction in cost compared to traditional multi-CPU platforms and increased performance, GPUs have become a common approach to parallelisation.
Particularly the use of NVIDIA GPUs and their programming language CUDA have become prevalent. 
CUDA today includes many GPU versions of common numerical libraries such as the linear algebra package cuBLAS, the Fourier transform library cuFFT and cuSPARSE which provides the programmer with many common solution methods for sparse matrices, discussion of these can be found on the CUDA documentation web-page~\cite{cudaDoc}. 

There is a large field of literature associated with the implementation finite-difference methods using CUDA, a few examples include~\cite{Micikevicius093dfinite,waveFD,elsStencil,CUDAthesis}.
This literature commonly explains how to approach the problem of implementing a finite difference scheme using CUDA but yet the authors provide no publicly available library or code with their papers that a reader readily use in their own project, thus requiring the reader to rewrite code that repeats work already done elsewhere.
Libraries providing PDE solvers and other stencil--based computations exist, such as \cite{libGeo} and indeed some approaches that can generate code for the programmer~\cite{autoGenZhang,autoGenHolewinski}, but these libraries and approaches can be limiting due to investment cost in learning essentially a full software package or new method.
Indeed the PETSc library~\cite{petscwebpage,petscuserref,petscefficient} which supports finite-difference methods also now provides a GPU implementation but this limits the program to be written mostly using that library’s API (thus limiting flexibility), and requires the programmer to also develop knowledge of cuTHRUST~\cite{cudaDoc} to implement the GPU aspects of the library effectively. 
It can also be noted that the PETSc web-page~\cite{petscwebpage} currently documents some difficulties associated with using GPUs effectively in PETSc. 
As such, we present cuSten as a computational framework complementary to PETSc, readily deployable by a programmer interested in Physics applications, with relatively low overhead in terms of learning to implement large complex libraries.

Common problems at development time include readjusting boundaries when changing finite difference schemes or ensuring the correct data has been loaded onto the GPU at the time of computation, both of these are dealt with by cuSten.
cuSten aims to overcome these difficulties along with addressing the problems with the above problems by providing a new software tool, introducing a simple set of four functions (three in many cases) for the programmer to implement their finite difference solver.
These functions are accessed much like cuBLAS or cuSPARSE giving freedom to the programmer to build the program as they choose but eliminating the need to worry about the finite difference implementation specifics.
This tool allows a programmer to simply input their desired finite-difference stencil and the direction in which it should be applied and then the rest of the implementation, including the domain decomposition, boundary positioning and data handling are wrapped into functions that are easily called.
This approach reduces the development time necessary for implementing new systems/solvers and provides a robust framework that does not involve a black-box-approach to the solution from the programmer.  Furthermore, the approach does not require a major overhead of time to invest in learning/implementing a new tool.

It is not intended that the code produced by this tool be the most efficient implementation of a given scheme versus a dedicated code for a specific problem, but it is intended that the development time of a code is drastically cut by removing the need for the programmer to do unnecessary work at development time.
We include a comprehensive example of the application of this tool in Section~\ref{sec:ADI}, here the ease of implementation of the cuSten library to solve the 2D Cahn--Hilliard equation is highlighted.
A benchmark of cuSten versus a serial implementation of the same is also included to highlight the improvements in performance due to parallelisation on the GPU. 
2D problems are the main focus of this new tool; 2D problems provide a test-bed for the development numerical algorithms which can then be extended to 3D, where debugging, testing, and validation are more time-consuming.  
The extension of the present method to 3D is discussed in Section~\ref{sec:con} below.
In terms of floating point precision this library focuses on the use of the double floating point type as in most application it is desirable to have 64 bit precision when solving PDEs, the source code is easily modified using a standard text editor with find and replace to change to other data types if so desired (this is discussed also in Section~\ref{sec:con} below).

In Section~\ref{sec:arch} we introduce the underlying architecture of the cuSten library, including how it uses streams and events for optimal memory management. 
Then in Sections~\ref{sec:fun} and~\ref{sec:ex} we talk the reader through the cuSten API along with examples and where to find all the source code within the library should they wish to edit it, the API is further explained in the Doxygen documentation included with the library itself.
Section~\ref{sec:ADI} presents the implementation of a full 2D Cahn--Hilliard solver along with a GPU versus CPU benchmark of the cuSten library and finally concluding discussions with potential future work are presented in Section~\ref{sec:con}.


\section{Software Architecture}
\label{sec:arch}
The library in this paper makes use of the CUDA programming language. 
For the sake of brevity we assume the reader is familiar the standard features of the language including kernels, shared memory etc.
The tool is built on two main sets of code, one handling the creation and destruction of the \codeword{cuSten\_t} data type which handles all of the programmer's inputs (found in \codeword{/src/struct}) and the other handling the compute kernels (found in \codeword{/src/kernels}). 

At the top level will be the main program solving whatever PDE is of concern to the programmer and the library is called through the header \codeword{cuSten.h}.
The programmer provides the necessary memory to the library using Unified Memory along with the stencil details, these will be detailed in Section~\ref{sec:fun}. 
Unified Memory was chosen as it simplifies the handling of memory in the library and interfacing with the rest of programmer's code. 
The ability to address data beyond the device memory limit is also useful in cases where not all the data required for a given program fits in device memory, the movement of memory on and off the device is handled by the cuSten library as explained in the following paragraph.

To take advantage of Unified Memory the library allows the programmer to divide their domain into `tiles' such that each tile will fit into the device RAM.
Each tile is a chunk of the total domain in the y direction to ensure the memory is contiguous.
The tiles are loaded onto the GPU in time for the kernel to be launched such that there are no GPU page faults.
The programmer also has the option to unload the tiles onto host RAM after the computation is completed on a given tile, this can be for IO or if the programmer needed to free device RAM for the next tile or a new task.
This system ensures that loading/unloading data and computation is implemented as a pipeline using separate streams for data loading/unloading and kernel launches ensuring that everything overlaps and ensuring that as little time is wasted retrieving memory over the PCIe bus which is a bottleneck to a memory bound program.
Finite difference programs, such as the ones discussed in this article, are typically memory bound as only a few computations are required per point in the array yet the memory overhead can be quite large when several variables need to have stencils applied to them. 
Events are used to ensure the data has been loaded prior to the launch of a kernel. 

The programmer has the choice of supplying a standard linear stencil or a function pointer with additional input coefficients to the library, examples of which are discussed in Sections~\ref{sec:weights} and~\ref{sec:pointer} respectively.
Within the compute kernel blocks of data with suitable boundary halos are loaded into shared memory.
The stencil or function is then applied to the block with each thread calculating the output for its position.
When this has completed the data is then output as blocks into the memory provided by the programmer for output, the same memory cannot be used for both as the blocks require overlapping data and thus cannot use already output values.

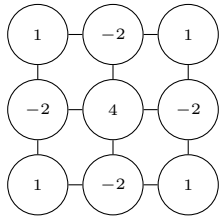
\begin{figure}
	\centering
\begin{tikzpicture}
  \stencilpt{-1,-1}{i-5}{$1$};
  \stencilpt{0,-1}{i-4}{$-2$};
  \stencilpt{1,-1}{i-3}{$1$};
  \stencilpt{-1,0}{i-2}{$-2$};
  \stencilpt{0,0}{i-1}{$4$};
  \stencilpt{1,0}{i}{$-2$};
  \stencilpt{-1,1}{i+1}{$1$};
  \stencilpt{0,1}{i+2}{$-2$};
  \stencilpt{1,1}{i+3}{$1$};
  \draw (i-5) -- (i-4)
        (i-4) -- (i-3)
        (i-5)   -- (i-2)
        (i-4) -- (i-1)
        (i-3) -- (i)
        (i-1) -- (i)
        (i)   -- (i+3)
        (i+1) -- (i+2)
        (i-2) -- (i-1)
        (i+3) -- (i+2)
        (i-1) -- (i+2)
        (i-2) -- (i+1);
\end{tikzpicture}
\caption{Typical stencil for a second order accurate cross derivative $\frac{\partial^4}{\partial x^2 \partial y^2}$.}
\label{fig:stencil}
\end{figure}


\section{Software API}
\label{sec:fun}
The programmer can use up to four functions for the application of any given finite difference stencil, in most cases only three are required.
The possible stencil directions include x, y and xy, where xy allows for cross derivatives which require that diagonal/off-diagonal information is available for the stencil to be completed, the stencil size is not limited in any direction and can be any desired shape, for example the stencil can be a $5 \times 3$ in dimension and use every point within that area.
Indeed the area for the stencil need not be centred at $(i, j)$ and it can extend in any direction more than another as necessary, this can be done be specifying non symmetric quantities for the number of points required left/right or top/bottom of $(i, j)$ in the stencil.
A typical stencil for a second order accurate cross derivative $\frac{\partial^4}{\partial x^2 \partial y^2}$ is shown in Figure~\ref{fig:stencil}, this stencil also appears in the linear biharmonic term for the Cahn--Hilliard solver presented in section~\ref{sec:adiApply}.

Each direction then comes with a periodic and non-periodic boundary option along with a choice between supplying just a set of weights (example in \ref{sec:weights}) which are applied linearly or a function pointer (examples in Section~\ref{sec:pointer} and~\ref{sec:adiApply}) that can be used to apply more sophisticated schemes.
In order to apply non--periodic boundary conditions the programmer will need to write their own boundary condition kernel, this was done to keep the library flexible to the programmer's desired numerical scheme which may require more sophisticated boundaries than simple Neumann/Dirichlet conditions. 
The cuSten library simply leaves the data in the boundary cells untouched when performing a non--periodic computation.
The naming convention for the functions available in the library is 
\[\codeword{cuSten[Create/Destroy/Swap/Compute]2D[X/Y/XY][p/np][BLANK/Fun]}\]
The descriptions for the options are as follows:

\codeword{Create}: This will take the programmer inputs such as the stencil size, weights, number of tiles to use etc. and return the cuSten\_t ready for use later in the code.

\codeword{Destroy}: This will undo everything in create, freeing pointers and streams etc. To be used when the programmer has finished using the current stencil, for example at the end of a program.

\codeword{Swap}: This will swap all relevant pointers, in other words swap the input and output data pointers around so the stencil can be applied to the updated stencil after time-stepping. The need for this function is generally dependent on the overall numerical scheme a programmer is using, it is not needed in all situations.

\codeword{Compute}: This will run the computation applying the stencil to the input data and outputting it to the appropriate output pointer.

\codeword{X}: Apply the stencil in the x direction.

\codeword{Y}: Apply the stencil in the y direction.

\codeword{XY}: Apply the stencil in the xy direction simultaneously (for situations with cross derivatives etc.). The library will account for corner halo data in this situation.

\codeword{p}: Apply the stencil with periodic boundary conditions.

\codeword{np}: Apply the stencil with non-periodic boundary conditions, this leaves suitable boundary cells untouched for the programmer to then apply their own boundary conditions.

\codeword{Fun}: Version of the function to be used if supplying a function pointer, otherwise leave blank.

The functions are then called in order of \codeword{Create}, \codeword{Compute}, \codeword{Swap} (if necessary) and then \codeword{Destroy} at the end of the program. 
Complete usage examples are found in the next section with further examples found in \codeword{examples/src}. 
The complete API can be found in the Doxygen documentation, see \codeword{README} on how to generate this.


\section{Examples}
\label{sec:ex}
In this section we provide an overview of using library.
We present three examples.  
The first is an implementation using linear stencil weights.  
The second involves a function pointer instead.  
The third example is at the level of a detailed physics problem (advection in Fluid Mechanics), and is included here to demonstrate to the user how to modify the source code as necessary.
These three examples (and more) can be found in \codeword{examples/src}. 
The \codeword{README} provides compilation details.
In all examples in the repository we take derivatives of various trigonometric functions as these are easy to benchmark against in periodic and non-periodic domains.

\subsection{Standard Weights}
\label{sec:weights}
We present here the example \codeword{2d_x_np.cu}, it is recommended to have this example open in a text editor to follow along. 
In this example we implement an 8th order accurate central difference approximation to the second derivative of $\sin(x)$ in the $x$ direction.
The domain has 1024 points in $x$ and 512 points in $y$, set by nx and ny respectively with the domain size lx set to $2 \pi$.

Unified memory is allocated with \codeword{dataOld} set to the input $\sin(x)$ and answer set to $-\sin(x)$, \codeword{dataNew} is zeroed to ensure correct output.
We choose to implement this scheme on compute device 0 by setting \codeword{deviceNum} and implement the scheme using a single tile, setting \codeword{numTiles} to 1.
The stencil is then implemented by setting the parameters \codeword{numSten}, \codeword{numStenLeft} and \codeword{numStenRight} along with providing an array of the stencil weights the same length as \codeword{numSten}.
\codeword{numSten} is the total number of points in the stencil, in this case 9, while \codeword{numStenLeft/Right} are the number of points in the left and right of the stencil, both 4 in this case. 
A \codeword{cuSten\_t} named \codeword{xDirCompute} is then declared and fed along with the above parameters into custenCreate2DXnp, this then equips \codeword{cuSten_t} with the necessary information. 
The ordering of parameters to be fed into \codeword{cuStenCreate2DXnp} can be found in both the Doxygen documentation and \codeword{cuSten/src/struct/cuSten_struct_functions.h}

The computation is  run using \codeword{cuStenCompute2DXnp(&xDirCompute, HOST)} where the \codeword{HOST} indicates we wish to load the data back to the host memory after the computation is completed, \codeword{DEVICE} if you wish to leave it in device memory.
Finally the result is output along with the expected answer to \codeword{stdout}, the 4 cells on either side in the x direction will be $0.0$ due to the boundary, these would then be set by the programmer using suitable boundary conditions in a full solver.
Then the \codeword{cuStenDestroy2DXnp} function is called to destroy the \codeword{cuSten\_t}.
Memory is then freed in the usual manner.

\subsection{Function Pointer}
\label{sec:pointer}
Now we present the function pointer version of the above example, named \codeword{2d_x_np_fun.cu}, again is is recommended to have a text editor open with the code to follow along.
Many of the parameters are the same as before except this time we remove the weights and replace them with coefficients that are then fed into the function pointer by the library.

The function pointer in this case implements a standard second-order accurate central-difference approximation to the second derivative of $\sin(x)$. 
We supply \codeword{numSten}, \codeword{numStenLeft} and \codeword{numStenRight} as before but now we also need \codeword{numCoe} to specify how many coefficients we need in our function pointer.

Our function pointer is of type \codeword{devArg1X}, where the 1 indicates how many input data sets are required. 
Each thread in a block will call the function and it returns the desired output value for that thread, each index in the array has one thread assigned to it.
The inputs are pointers to the input data, the coefficients and the index location in the stencil
\[
\codeword{CentralDifference(double* data, double* coe, int loc)}
\]
The central-difference scheme is implemented in a standard way with indexing done relative to loc, the coefficient in this case is set to $1.0 / \Delta x ^2$ as is standard.
A key point to notice, is that the programmer must allocate memory for the function pointer on the device, this can be seen on line 131 and 132 of the example code prior to calling the \codeword{Create} function.

The rest of the access to the API is then the same as before except some of the inputs change and there is 
a \codeword{Fun} at the end of each function name, for example \codeword{cuStenCreate2DXnpFun}. 
We will see later in Section~\ref{sec:ADI} how function pointers provide us with a powerful tool to apply stencils to non-linear quantities, in particular we will see this with the cubic term of the Cahn--Hilliard equation to which we wish to apply a Laplacian. 

\subsection{Advection}
The library also comes with an extra variant of the above functions \codeword{2d_xyADVWENO_p} in which a 2D periodic advection WENO scheme has been implemented by modifying the \codeword{2DXYp} source code. 
This is included as an example to show the user how to modify the source code as necessary to more specific needs or in situations where the function pointer may not meet requirements, for example in this situation where extra data needed to be input in the form of $u$ and $v$ velocities. 
The files can be found in the cuSten/src folder with how its called in \codeword{examples/src/2d_xyWENOADV_p.cu}.

A brief overview of the modifications made to the \codeword{2DXYp} code are as follows:

\begin{itemize}
\item The stencil dimensions are now set automatically when the creation function is called.
\item The $u$ and $v$ velocities were linked to the cuSten type with appropriate tiling.
\item Additional asynchronous memory copies were included in the memory loading portion of the code to ensure the velocities are present on the device at the required time.
\item The corner data copying to shared memory blocks was removed from the kernel as it is no longer required.
\item The standard stencil compute was removed and replaced with a device function call to a WENO solver, details of the solver can be found in~\cite{osher2006level}.
\end{itemize}


\section{cuCahnPentADI}
\label{sec:ADI}
In this section we show how the cuSten library can be used as part of a larger program that the authors developed using the cuPentBatch~\cite{cuPent} solver, a batched pentadiagonal matrix solver.
We also provide a benchmark at the end of the section to show how cuSten performs versus a serial implementation.
The equation we wished to solve was the 2D Cahn--Hilliard equation. 
The Cahn--Hilliard equation models phase separation in a binary liquid: when a binary fluid in which both components are initially well mixed undergoes rapid cooling below a critical temperature, both phases spontaneously separate to form regions rich in the fluid's component parts. 
The regions expand over time in a phenomenon known as coarsening~\cite{CH_orig}.   
The equation is extremely well studied and is a popular model in polymer physics and interfacial flows.

In the mathematical framework, a single scalar concentration field $C(\vecx,t)$ characterizes the binary mixture.  
As such, a concentration level $C=\pm 1$ indicates phase separation of the mixture into one or other of its component parts, while $C=0$ denotes a perfectly mixed state.  
The free energy for the mixture can be modeled as 
\begin{equation}
F[C]=\int_\Omega \left[\frac{1}{4}(C^2 -1)^2 +\frac{1}{2}\gamma|\nabla C|^2\right] \mathd^\mydim x
\end{equation}
where the first term promotes de-mixing and the second term smooths out sharp gradients in transition zones between de-mixed regions; also, $\gamma$ is a positive constant, $\Omega$ is the container where the binary fluid resides, and $\mydim$ is the dimension of the space. 
The twin constraints of mass conservation and energy minimization suggest a gradient-flow dynamics for the evolution of the concentration: 
\begin{equation}
\frac{\partial C}{\partial t} = \nabla\cdot\left[D(C)\nabla \frac{\delta F}{\delta C}\right] 
\end{equation}
where $\delta F/\delta C$ denotes the functional derivative of the free energy and $D(C)\geq 0$ is the mobility function, assumed for simplicity in this work to be a positive constant.  
As such, the basic model equation reads
\begin{subequations}
\begin{equation}
\frac{\partial C}{\partial t}=D\nabla^2\left(C^3-C-\gamma\nabla^2 C\right),\qquad \vecx\in \Omega,\qquad t>0.
\label{eq:ch_basic}
\end{equation}
The initial condition is given as
\begin{equation}
C(\vecx,t=0)=f(\vecx),\qquad \vecx\in \overline{\Omega}.
\end{equation}%
\label{eq:ch_all}%
\end{subequations}%
\subsection{Discretisation}
For simplicity, we focus on the case where $\Omega=(0,2 \pi)^\mydim$, with periodic boundary conditions applied in each of the $\mydim$ spatial dimensions.  The method of solution we choose is based on the ADI method presented in~\cite{stableHyper} for the linear hyperdiffusion equation -- we extend that scheme here and apply it to the non-linear Cahn--Hilliard equation as follows:

\begin{subequations}
\begin{align}
{\bf{L_x}}w = -\frac{2}{3}(C^{n} - C^{n-1}) - \frac{2}{3} \Delta t \nabla^4 \bar{C}^{n+1} + \frac{2}{3}D \Delta t \nabla^2(C^3 - C)^n \\
{\bf{L_y}}v = w \\
C^{n + 1} = \bar{C}^{n+1} + v,
\end{align}%
\label{eq:ch_adi}%
\end{subequations}%
Where ${\bf{L_x}} = {\bf{I}} + \frac{2}{3} D \gamma \Delta t \partial_{xxxx}$ and similarly for ${\bf{L_y}}$.
In Equation~\eqref{eq:ch_adi} each one of the matrix inversions is solved using cuPentBatch as per the method presented in \cite{cuPent} and we transpose the matrix when changing from the x direction to y direction sweep to ensure the data is in the proper interleaved format.
To deal with the periodic element of the inversion the method is the same as in Reference~\cite{cuPent,navon_pent}. 
To recover the initial $n-1$ time step we simply set this to the initial condition and appropiately update and time steps there after. 
The derivatives are discretised using standard second order accurate central differences.
\begin{subequations}
\begin{eqnarray}
\frac{\partial^2 \phi_i}{\partial x^2} = \frac{\phi_{i+1}-2\phi_i + \phi_{i-1}}{\Delta x^2},\\
\frac{\partial^4 \phi_i}{\partial x^4} = \frac{\phi_{i+2}-4\phi_{i+1}+6\phi_i - 4\phi_{i-1}+\phi_{i-2}}{\Delta x^4},
\end{eqnarray}%
\end{subequations}%
and $\Delta x=\Delta y$ for the a uniform grid.

\subsection{Application of cuSten}
\label{sec:adiApply}
The code for the example can be found with the repository in the \codeword{cuCahnPentADI} folder, supplied also in this folder is a \codeword{Makefile} to compile the files and a Python script to analyse the results which we present in Section~\ref{sec:results}.
cuSten is applied for all of the finite-difference elements of the code excluding the matrix inversion where we use cuPentBatch. 
Between lines 148 and 190 we can see an application of a more sophisticated function pointer than presented previously in Section~\ref{sec:pointer}, here we apply the Laplacian to the right-hand-side (RHS) non-linear term $C^3 - C$.
The coefficients are declared between lines 481 and 516, the non--linear term is a $5 \times 5$ stencil.
This shows a clear example of ease of use of the function pointers and the easy swap in/out of values. 
Note how the indexing starts from the top left of the stencil and sweeps left to right in i, row by row in j for indexing. 

The linear terms for the RHS are implemented using standard weighted schemes, the scheme uses a $5 \times 5$ stencil, this and the non--linear term highlight one of the key features of the library with the easy change in stencil size and the boundaries are dealt with automatically (in this case periodic).
The additional static functions at the start of the file apply the time stepping parts of the algorithm and combination of terms to set the full RHS. 
Output is done using the standard HDF5 library, this is required for the cuPentBatchADI program but not the cuSten library itself.

\begin{figure}[H]
	\centering
		\includegraphics[width=0.7\textwidth]{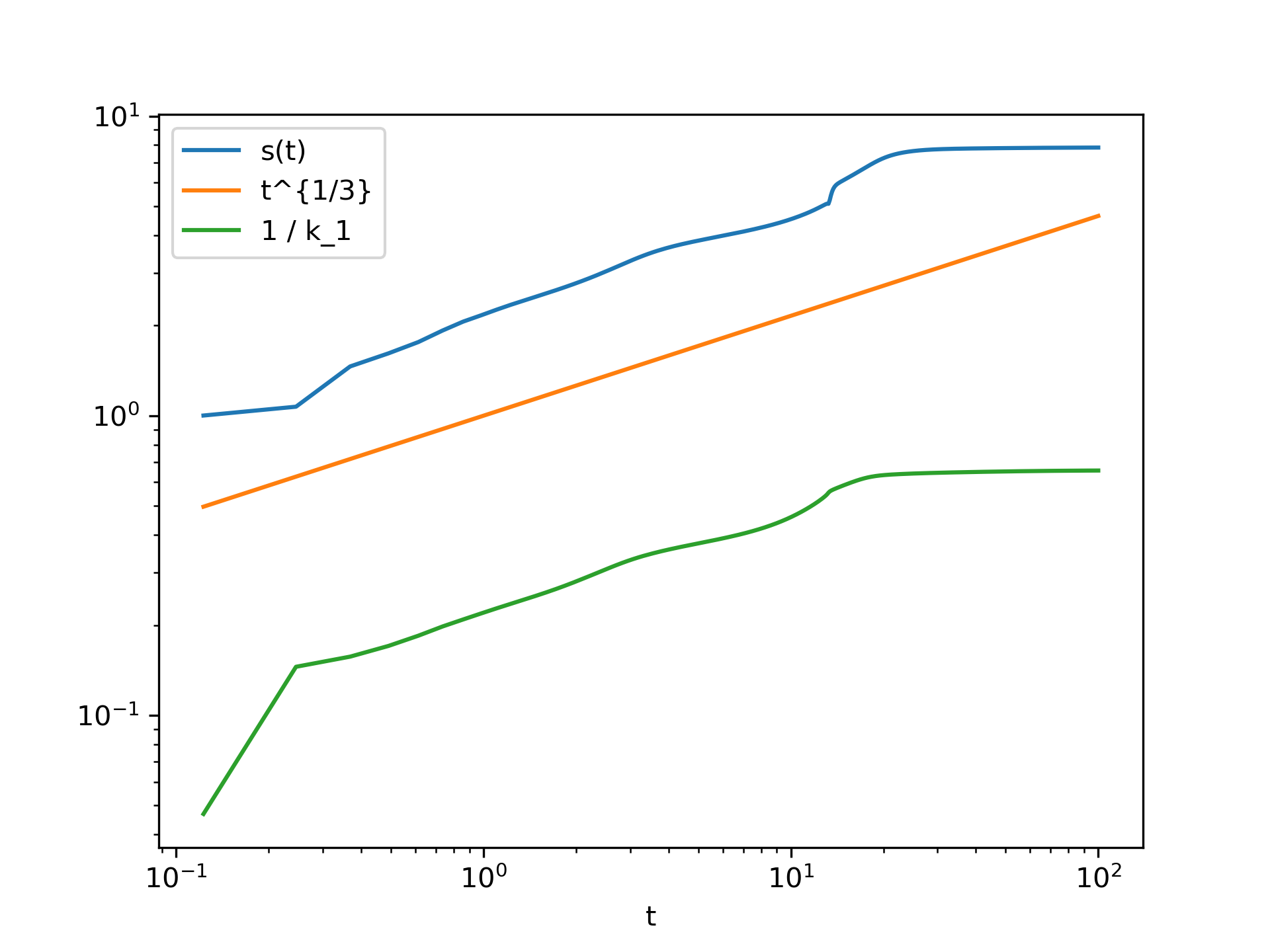}
		\caption{Plot showing $s(t)$ and $k_1$ as functions of $t$. We can see the clear $t^{1/3}$ behaviour as expected in each.}
	\label{fig:scaling}
\end{figure} 

\subsection{Numerical Results}
\label{sec:results}
\begin{figure}[H]
	\centering
	\subfigure[$t = 0.7363107782$]{\includegraphics[width=0.45\textwidth]{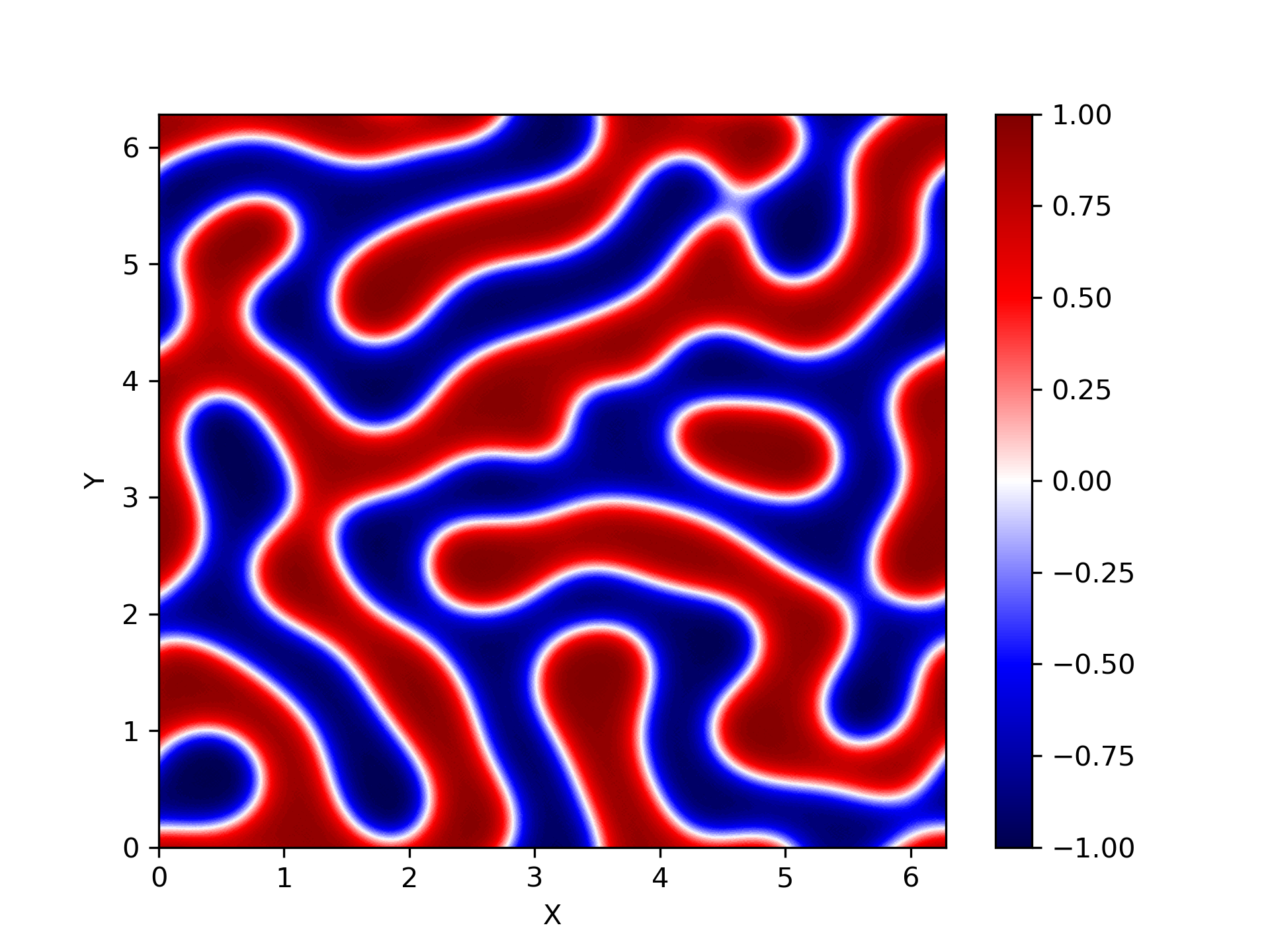}}
	\subfigure[$t = 3.1906800388$]{\includegraphics[width=0.45\textwidth]{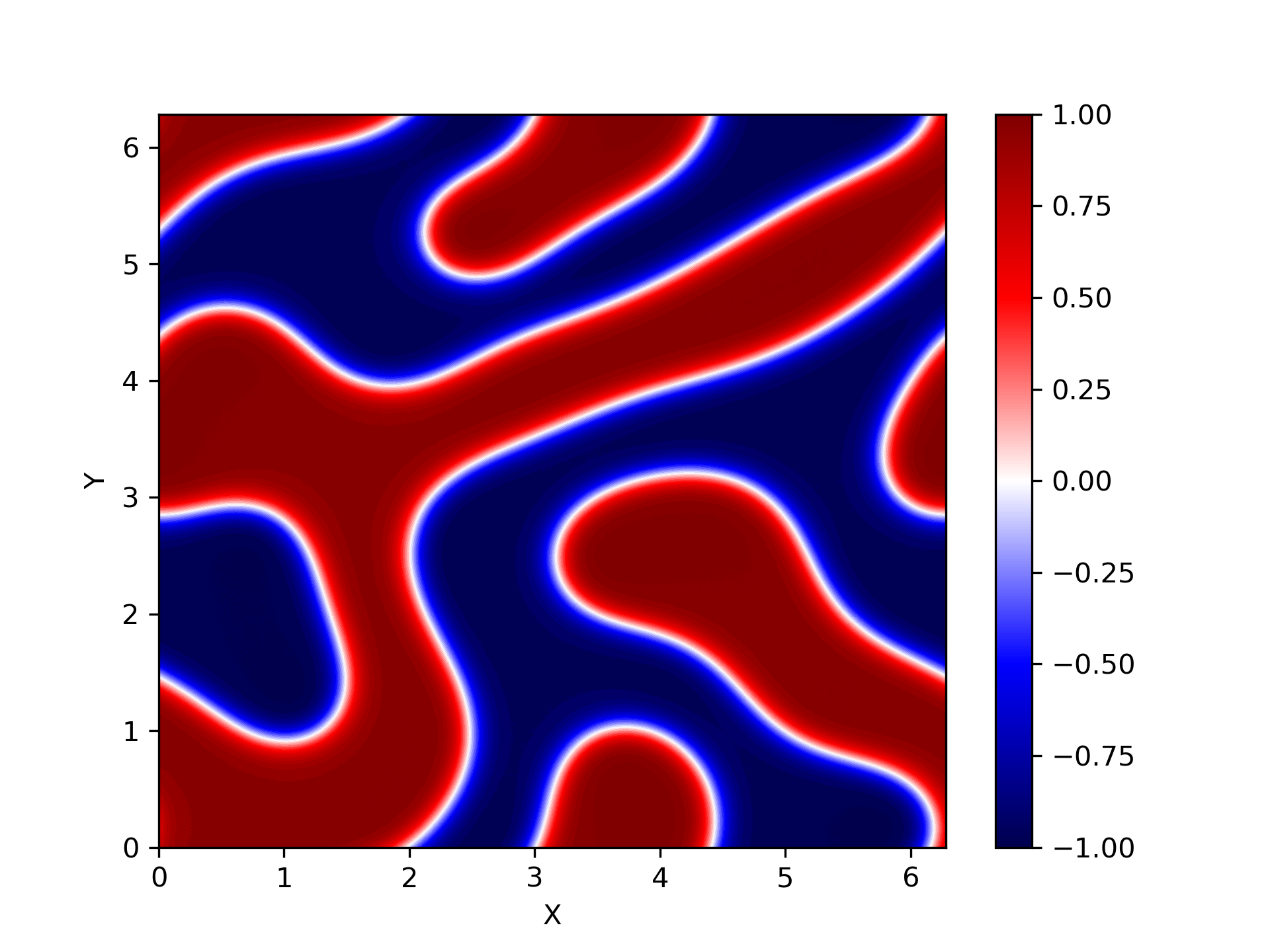}}
	\subfigure[$t = 12.3945647661$]{\includegraphics[width=0.45\textwidth]{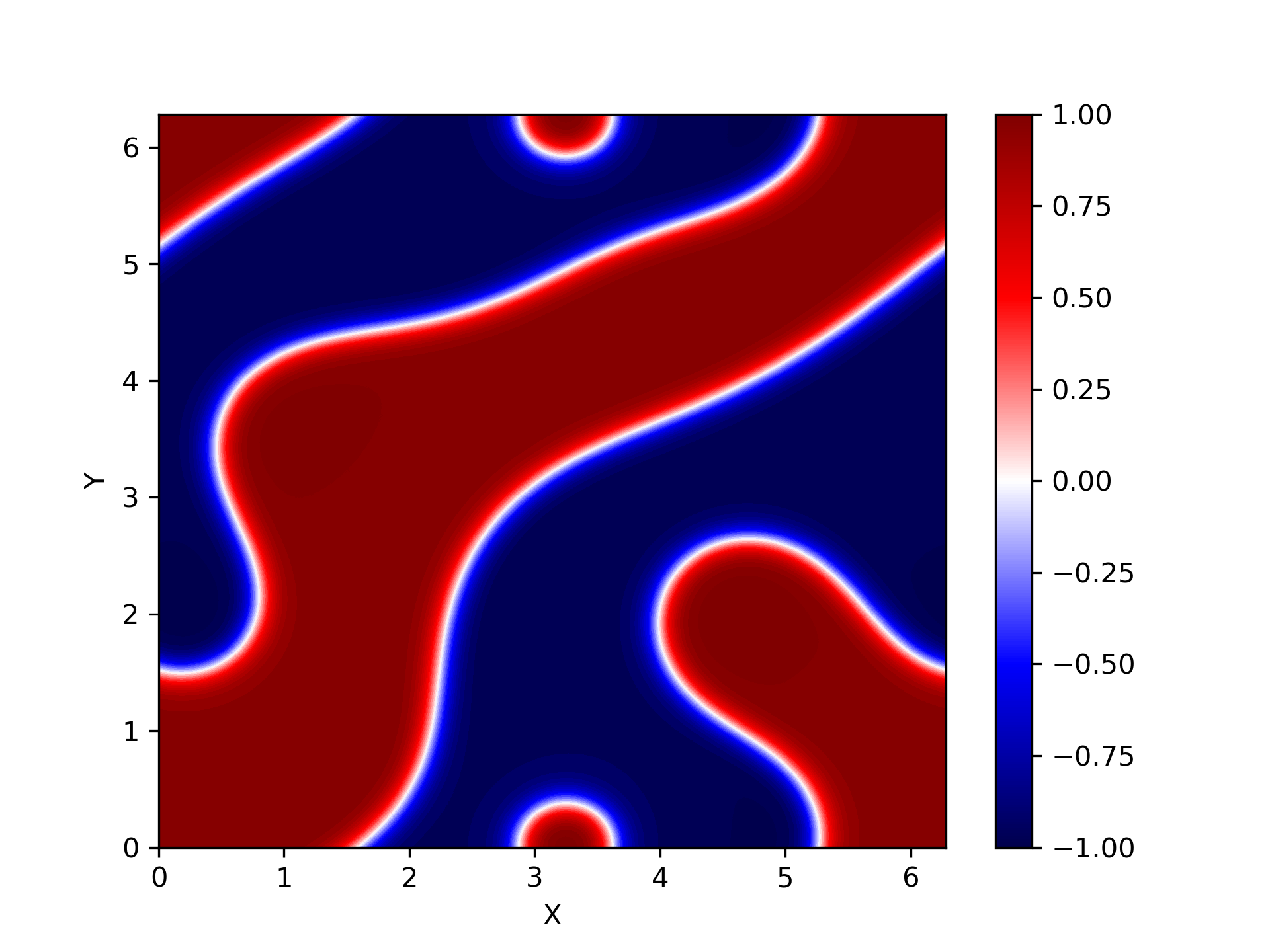}}
	\subfigure[$t = 73.7537962816$]{\includegraphics[width=0.45\textwidth]{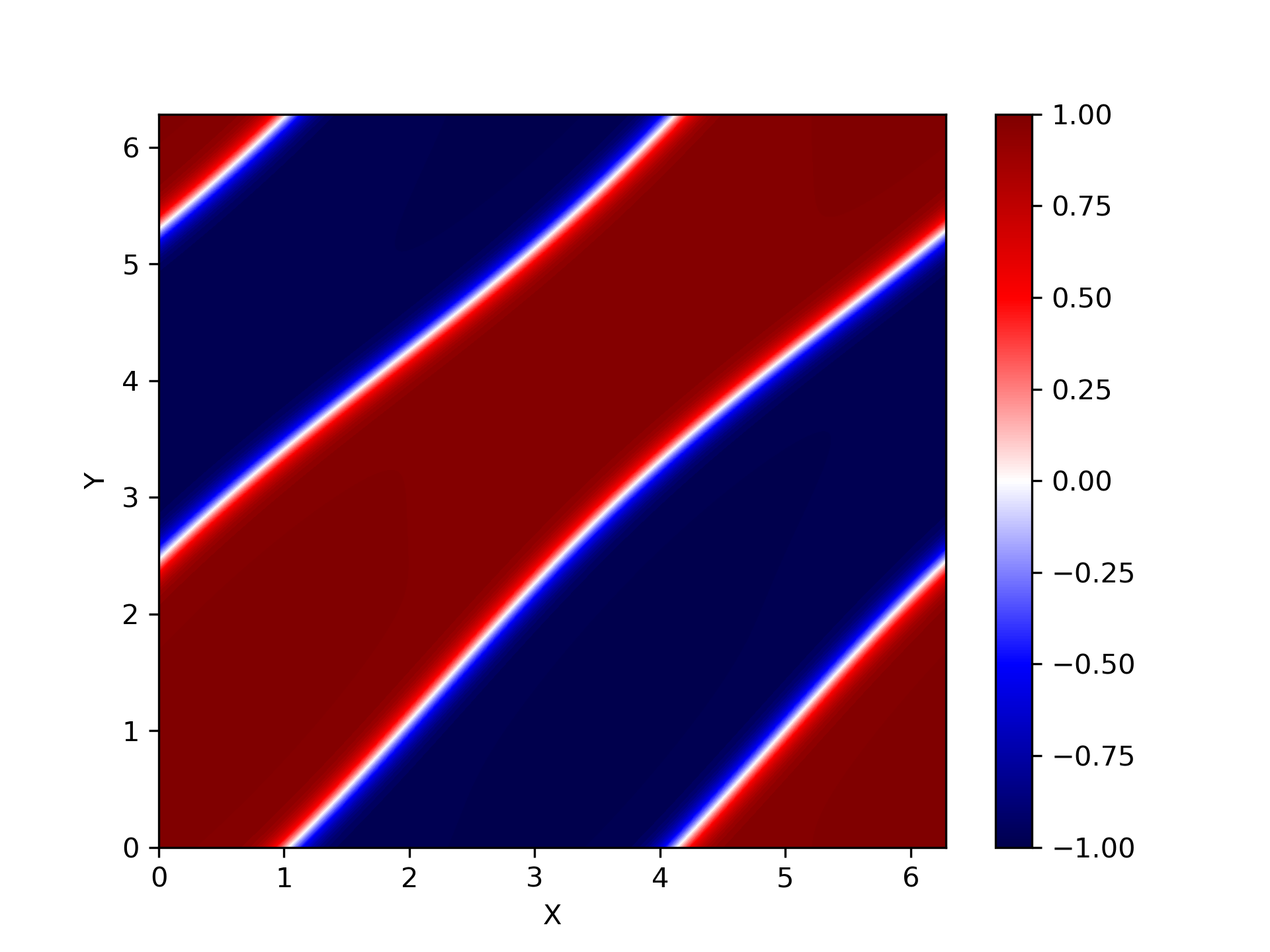}}
	\caption{Contour plot of solution at various times}
\label{fig:contour_all}
\end{figure}

In order to analyse the performance of the code we use two standard tests to quantify the coarsening rate \cite{LennonAurore}. 
First we have the quantity $s(t)$ which can be defined as 
\begin{equation}
s(t) = \frac{1}{1 - \langle C^2\rangle}
\end{equation} 
Where $\langle \cdot\rangle$ denotes the spatial average, which we calculate by a simple integration over the domain using Simpsons's rule.
Secondly we plot $1 / k_1(t)$, which also captures the growth in length scales, where $k_1$ can be defined as 
\begin{equation}
k_1(t) = \frac{\int d^nk |\hat{C}|^2}{\int d^nk |{\bf{k}}|^{-1} |\hat{C}|^2}
\end{equation}
with the hat denoting the Fourier Transform. 
We run the simulation to a final time $T=100$ with $n_x = n_y = 512$ points, the time--step size is set at $\Delta t = 0.1 \Delta x$.
The initial conditions are a random uniform distribution of values between $-0.1$ and $0.1$, we have set the coefficients $D$ and $\gamma$ to $1.0$ and $0.01$ respectively.  
The initial condition is chosen to mimic a `deep quench', where the system is cooled suddenly below the critical temperature, which allows for phase separation to occur spontaneously~\cite{Zhu_numerics}.
The quantities $s(t)$ and $1/k_1(t)$ are plotted in Figure~\ref{fig:scaling} as a function of $t$ with a reference line of $t^{1/3}$ included as both should scale proportionally to this.
We can see clear match between our two quantities and $t^{1/3}$. 
Finite-size effects spoil the comparison between numerics and theory towards the end of the computation, as by that time the $(C=\pm 1)$-regions fill out the computational domain.  
Figure~\ref{fig:contour_all} to illustrate the behaviour of the solution in space and time: the system clearly evolves into extended regions where $C=\pm 1$, which grow over time, consistent with Figure~\ref{fig:scaling} and the established theory~\cite{LS}, in particular we can see the development of the finite size effects.

\subsection{Benchmark of cuSten}
In this section we benchmark the cuPentCahnADI program, which uses cuSten and cuPentBatch, against a serial version of the program running on a CPU.
The GPU used in this benchmark is an NVIDIA Titan X Pascal and the CPU is an Intel i7--6850K which has 6 hyper-threaded cores operating at $3.6 GHz$.
The benchmark is performed by measuring the time to time--step the simulation to a final time of $T = 10$, scaling $N$ where $N \times N$ is the total size of the domain.
All start--up overheads and program--finish overheads are excluded from the timing, this is to ensure a fair benchmark of only the numerical computation, $T = 10$ was chosen to ensure any effects of background processes due to the operating system are averaged out.
No IO steps were included in either code.
The same parameters and initial conditions were used as in the previous section, the serial code and version of cuPentCahnADI which outputs times rather than simulation data can be found in the folder \codeword{cuPentSpeedUp}.

\begin{figure}[H]
  \centering
    \includegraphics[width=0.8\textwidth]{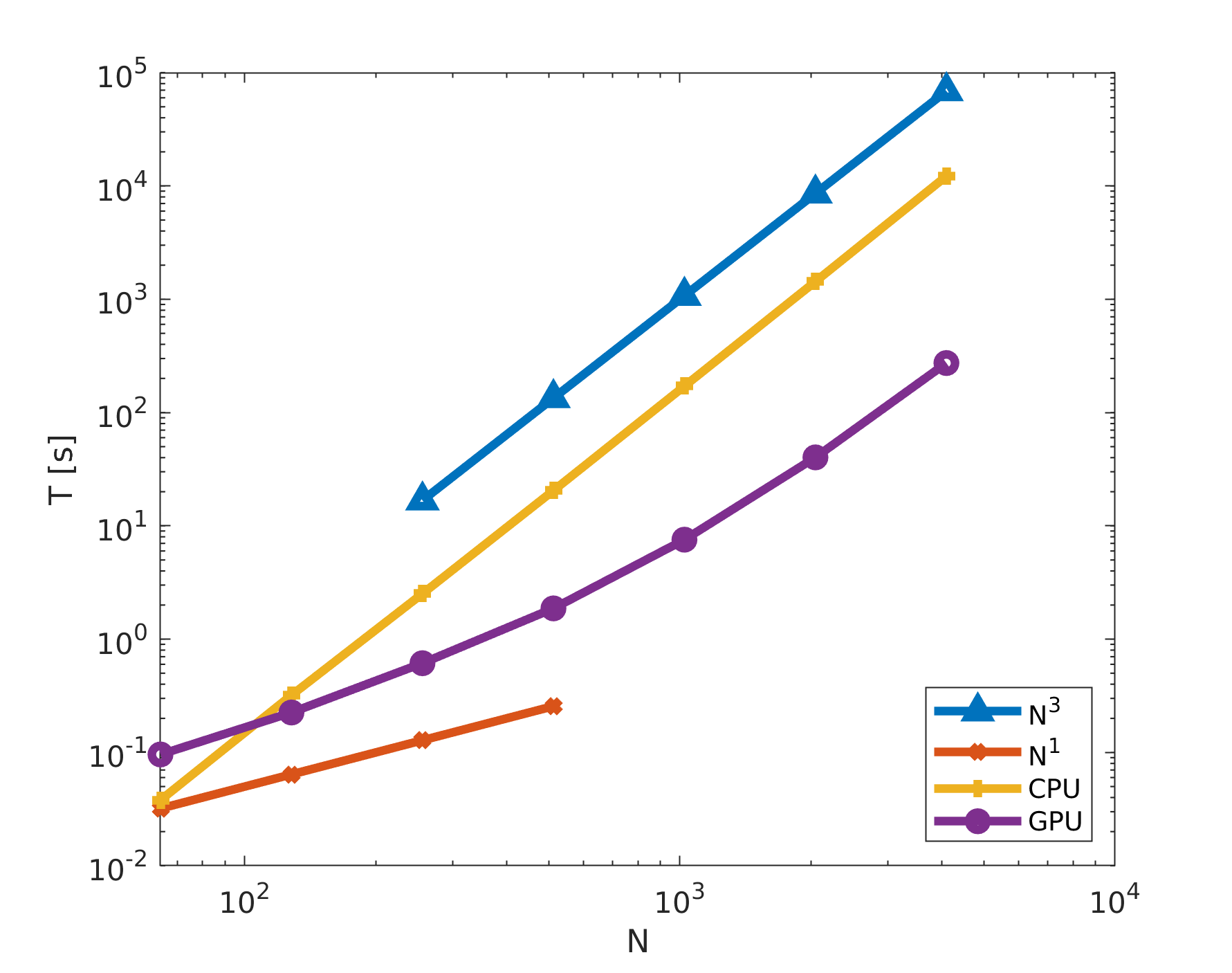}
    \caption{Plot showing how the CPU and GPU times scale with $N$.}
  \label{fig:timeScale}
\end{figure}

In Figure~\ref{fig:timeScale} we present the scaling in time as a function of $N$ for the serial and GPU codes, superimposed are the lines for $N^2$ and $N^3$ for comparison.
It can be seen clearly from these plots that the CPU code scales in time as $N^3$ while the GPU code initially scales with $N^2$ increasing to $N^3$ as $N$ increases.
This can be attributed to the presence of Amdahl's law, as $N$ increases more and more the serial computation required per core of the GPU will begin to dominate and the parallelisation speed-up levels out.

\begin{figure}[H]
  \centering
    \includegraphics[width=0.8\textwidth]{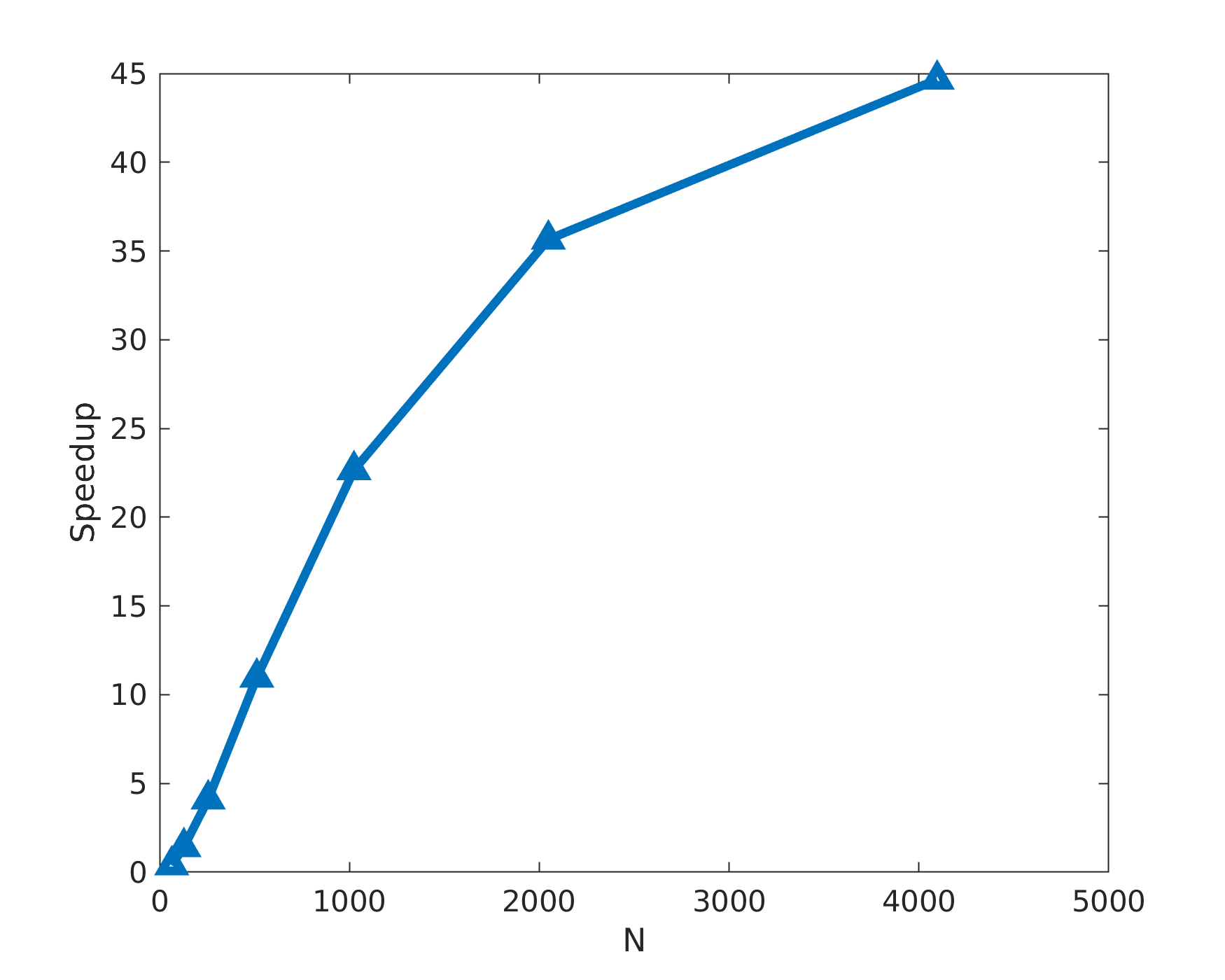}
    \caption{Plot showing speed-up of the GPU code versus the CPU code as a function of $N$.}
  \label{fig:speedup}
\end{figure}

Further evidence of this can be seen in Figure~\ref{fig:speedup} as the curve begins to level off for $N$ large. 
In this plot though we can see the clear advantage of parallelising this 2D solver on a GPU versus the serial CPU code, the speed-up is on the $O(10)$ for all reasonable grid resolutions, indeed the speed-up gets to $40\times$ faster for large $N$, a significant performance increase.
This performance would be increased further on newer GPUs such as the V100.
Thus significant performance can be gotten by using a GPU with the cuSten library for 2D computations.
The advantages of GPUs for the speed-up of batches of 1D problems has already been discussed in~\cite{cuPent}, the results presented in that paper used an earlier version of the cuSten library.

\section{Discussion and Conclusions}
\label{sec:con}
\subsection{Possible Future Extensions to the Library}
As previously mentioned the current library is limited to 2D uniform grids with double precision. 
Future areas of expansion could include moving the current library functions into C++ templates, this would allow for easier generalisation to other data types without the current need for find and replace to be done manually. 
Expansion to 3D and non-uniform grids is less trivial. 
3D would require a different approach to loading data than currently implemented as data will not be contiguous in RAM in the z direction, a more sophisticated loading scheme with pointers would be required.
For non-uniform grids additional data would need to be loaded into memory, it is likely in this situation that a hybrid of modifying the code such as in the WENO example to have extra data available ($u$ and $v$ velocities in the case of WENO, coordinate transformations in the case of a non-uniform grid) and using function pointers would be the best approach to make to the existing source. 

\subsection{MPI}
The design of the library lends itself to an MPI domain decomposition to be used in a hybrid code with the cuSten library.
Each MPI process could be assigned to a GPU using the deviceNum parameter, then the user would apply the non periodic versions of the stencils along with using MPI to swap the boundary halos.
Memory exchange is simplified in MPI due to the use of Unified Memory, the required data will be copied directly between GPU devices. 
This allows for the application of this library in much larger solvers which require more than just a single GPU.

\subsection{Concluding Remarks}
In this paper we have shown how cuSten can be used to simplify the implementation of finite difference programs in CUDA compared with other state of the art libraries such as PETSc.
cuSten has a lightweight interface with a minimal learning curve required to implement the functions as part of a wider project. 
The library has been benchmarked against a serial code using a Cahn--Hilliard solver and numerous examples are provided to show potential users how to use the functionality provided. 
It has wide ranging applications in finite-difference solver development and in further areas requiring stencil--based operations such as image processing and optimisation problems. 

\section*{Acknowledgements}
\label{}
Andrew Gloster acknowledges funding received from the UCD Research Demonstratorship.
All authors gratefully acknowledge the support of NVIDIA Corporation with the donation of the Titan X Pascal GPUs used for this research.
The authors also acknowledge the referees of this paper who provided insightful feedback with good suggestions to improve both the paper and cuSten library.


\section{Bibliography}
\bibliographystyle{unsrt}
\bibliography{cuStenBiblio}

\begin{thebibliography}{10}

\bibitem{doering1995applied}
Charles~R Doering and John~D Gibbon.
\newblock {\em Applied analysis of the Navier-Stokes equations}, volume~12.
\newblock Cambridge University Press, 1995.

\bibitem{osher2006level}
Stanley Osher and Ronald Fedkiw.
\newblock {\em Level set methods and dynamic implicit surfaces}, volume 153.
\newblock Springer Science \& Business Media, 2006.

\bibitem{hesthaven2018numerical}
Jan~S Hesthaven.
\newblock {\em Numerical methods for conservation laws: From analysis to
  algorithms}, volume~18.
\newblock SIAM, 2018.

\bibitem{wilmott_howison_dewynne_1995}
Paul Wilmott, Sam Howison, and Jeff Dewynne.
\newblock {\em The Mathematics of Financial Derivatives: A Student
  Introduction}.
\newblock Cambridge University Press, 1995.

\bibitem{whitham2011linear}
Gerald~Beresford Whitham.
\newblock {\em Linear and nonlinear waves}, volume~42.
\newblock John Wiley \& Sons, 2011.

\bibitem{cudaDoc}
NVIDIA.
\newblock Cuda toolkit documentation.
\newblock \url{https://docs.nvidia.com/cuda/}, 2019.

\bibitem{Micikevicius093dfinite}
Paulius Micikevicius.
\newblock 3d finite difference computation on gpus using cuda.
\newblock In {\em Proceedings of 2nd Workshop on General Purpose Processing on
  Graphics Processing Units, ACM International Conference Proceeding Series},
  pages 79--84. ACM, 2009.

\bibitem{waveFD}
David Michéa and Dimitri Komatitsch.
\newblock Accelerating a three-dimensional finite-difference wave propagation
  code using gpu graphics cards.
\newblock {\em Geophysical Journal International}, 182(1):389--402, 2010.

\bibitem{elsStencil}
Andreas Schäfer and Dietmar Fey.
\newblock High performance stencil code algorithms for gpgpus.
\newblock {\em Procedia Computer Science}, 4:2027 -- 2036, 2011.
\newblock Proceedings of the International Conference on Computational Science,
  ICCS 2011.

\bibitem{CUDAthesis}
Dheevatsa Mudigere.
\newblock Data access optimized applications on the gpu using nvidia cuda.
\newblock {\em Master's Thesis, Technische Universit{\"a}t M{\"u}nchen}, 2009.

\bibitem{libGeo}
Andreas Sch\"{a}fer and Dietmar Fey.
\newblock Libgeodecomp: A grid-enabled library for geometric decomposition
  codes.
\newblock In {\em Proceedings of the 15th European PVM/MPI Users' Group Meeting
  on Recent Advances in Parallel Virtual Machine and Message Passing
  Interface}, pages 285--294, Berlin, Heidelberg, 2008. Springer-Verlag.

\bibitem{autoGenZhang}
Yongpeng Zhang and Frank Mueller.
\newblock Auto-generation and auto-tuning of 3d stencil codes on gpu clusters.
\newblock In {\em Proceedings of the Tenth International Symposium on Code
  Generation and Optimization}, CGO '12, pages 155--164, New York, NY, USA,
  2012. ACM.

\bibitem{autoGenHolewinski}
Justin Holewinski, Louis-No\"{e}l Pouchet, and P.~Sadayappan.
\newblock High-performance code generation for stencil computations on gpu
  architectures.
\newblock In {\em Proceedings of the 26th ACM International Conference on
  Supercomputing}, ICS '12, pages 311--320, New York, NY, USA, 2012. ACM.

\bibitem{petscwebpage}
Satish Balay, Shrirang Abhyankar, Mark~F. Adams, Jed Brown, Peter Brune, Kris
  Buschelman, Lisandro Dalcin, Alp Dener, Victor Eijkhout, William~D. Gropp,
  Dmitry Karpeyev, Dinesh Kaushik, Matthew~G. Knepley, Dave~A. May,
  Lois~Curfman McInnes, Richard~Tran Mills, Todd Munson, Karl Rupp, Patrick
  Sanan, Barry~F. Smith, Stefano Zampini, Hong Zhang, and Hong Zhang.
\newblock {PETS}c {W}eb page.
\newblock \url{https://www.mcs.anl.gov/petsc}, 2019.

\bibitem{petscuserref}
Satish Balay, Shrirang Abhyankar, Mark~F. Adams, Jed Brown, Peter Brune, Kris
  Buschelman, Lisandro Dalcin, Alp Dener, Victor Eijkhout, William~D. Gropp,
  Dmitry Karpeyev, Dinesh Kaushik, Matthew~G. Knepley, Dave~A. May,
  Lois~Curfman McInnes, Richard~Tran Mills, Todd Munson, Karl Rupp, Patrick
  Sanan, Barry~F. Smith, Stefano Zampini, Hong Zhang, and Hong Zhang.
\newblock {PETS}c users manual.
\newblock Technical Report ANL-95/11 - Revision 3.11, Argonne National
  Laboratory, 2019.

\bibitem{petscefficient}
Satish Balay, William~D. Gropp, Lois~Curfman McInnes, and Barry~F. Smith.
\newblock Efficient management of parallelism in object oriented numerical
  software libraries.
\newblock In E.~Arge, A.~M. Bruaset, and H.~P. Langtangen, editors, {\em Modern
  Software Tools in Scientific Computing}, pages 163--202. Birkh{\"{a}}user
  Press, 1997.

\bibitem{cuPent}
Andrew Gloster, Lennon~\'O N\'araigh, and Khang~Ee Pang.
\newblock cupentbatch a batched pentadiagonal solver for nvidia gpus.
\newblock {\em Computer Physics Communications}, 241:113 -- 121, 2019.

\bibitem{CH_orig}
J.~W. Cahn and J.~E. Hilliard.
\newblock Free energy of a nonuniform system. i. interfacial energy.
\newblock {\em J. Chem. Phys}, 28:258--267, 1957.

\bibitem{stableHyper}
T.P. Witelski and M.~Bowen.
\newblock Adi schemes for higher-order nonlinear diffusion equations.
\newblock {\em Applied Numerical Mathematics}, 45(2):331 -- 351, 2003.

\bibitem{navon_pent}
I.~M. Navon.
\newblock Pent: A periodic pentadiagonal systems solver.
\newblock {\em Communications in Applied Numerical Methods}, 3(1):63--69, 1987.

\bibitem{LennonAurore}
Aurore Naso and Lennon~\'O N\'araigh.
\newblock A flow-pattern map for phase separation using the navier stokes cahn
  hilliard model.
\newblock {\em European Journal of Mechanics - B/Fluids}, 72:576 -- 585, 2018.

\bibitem{Zhu_numerics}
Jingzhi Zhu, Long-Qing Chen, Jie Shen, and Veena Tikare.
\newblock Coarsening kinetics from a variable-mobility cahn-hilliard equation:
  Application of a semi-implicit fourier spectral method.
\newblock {\em Physical Review E}, 60(4):3564, 1999.

\bibitem{LS}
I.~M. Lifshitz and V.~V. Slyozov.
\newblock The kinetics of precipitation from supersaturated solid solutions.
\newblock {\em J. Chem. Phys. Solids}, 19:35--50, 1961.

\end{thebibliography}


\end{document}